# Residence Time Distribution Analysis of Aerosol Transport and Associated Healthcare Worker Exposure in a Mock Hospital Isolation Room via Computational Fluid Dynamics

Anthony J. Perez[1], Juan Penaloza-Gutierrez[2], Tauhidur Rahman[3], Andrés E. Tejada-Martínez[1*]

[1][Civil and Environmental Engineering, University of South Florida]

[2][Mechanical Engineering, University of South Florida]

[3][Halıcıoğlu Data Science Institute, University of California San Diego]

**\*CONTACT**  Andrés E. Tejada-Martínez, [aetejada@usf.edu] | Department of Civil and Environmental Engineering, University of South Florida, 4202 E Fowler Avenue, ENG 030, Tampa, Florida, 33620.

**Keywords:** mean residence time, turbulent diffusion, dead zones, short-circuiting

# ABSTRACT

The transport of aerosol discharge in the form of a passive scalar or tracer discharged from a single cough of a patient in a ventilated mock hospital isolation room is investigated via computational fluid dynamics (CFD). Healthcare worker (HCW) exposure to the aerosol is assessed through residence time analysis of the aerosol transported through the imperfect mixing conditions in the room. Flow features responsible for imperfect mixing, including short-circuiting or channeling between the patient and exhaust air vent (which leads to rapid expulsion of aerosols from the room), dead zones or re-circulation flow regions in the room, and the turbulent diffusion or spreading of aerosol across the room, are shown to play important factors determining the HCW exposure to the aerosol. The importance of each of these factors varies depending on the ventilation rate (ACH) and the placement of the exhaust air vent relative to the patient. For example, reducing ACH from 12 to 6 diminishes the importance of these flow features and the aerosol transport may be approximately modeled through the classical perfectly mixed assumption. At ACH = 12, especially when the ceiling exhaust is placed above the patient and the HCW, short-circuiting is the dominant feature in determining HCW exposure. But when the ceiling exhaust is placed away from the patient and HCW, the short-circuiting is weakened and the influence of dead zones, which trap aerosol, and turbulent diffusion, which allow the aerosol to escape, becomes more important. It is shown that the importance of these flow features and the resulting impact on HCW exposure can be quantified in terms of residence time distribution (RTD) metrics such as mean residence time and cumulative RTD. The results suggest that residence time analysis is a useful technique to be employed via CFD simulations and/or physical experiments when designing a hospital isolation room and assessing HCW exposure to aerosols.

# INTRODUCTION

To address the urgent needs of the healthcare community brought on by the COVID-19 pandemic, today's researchers have been pressed with employing data and science-driven approaches to deliver practical solutions. Over the last decades, rapid increase in computational power has allowed the use of computational fluid dynamics (CFD) to investigate aerosol dispersion in healthcare settings (Aliabadi et al. 2011), motivated by aerosol transmissibility of viruses, especially after the 2002-2004 SARS outbreak, the 2009 swine flu pandemic (Qian et al. 2009; Zhu et al. 2012), and the COVID-19 pandemic (Sheikhnedjad et al. 2022). Although CFD has high computational requirements, the relatively small dimensions of hospital isolation rooms have allowed it to provide good understanding of the dynamics of aerosol spreading in these settings (Memarzadeh and Xhu 2012; Thatiparti et al. 2016a; Thatiparti et al. 2016b; Cho 2019). CFD offers time-dependent, three-dimensional prediction of fluid flows, resulting in a clear understanding of flow features. When combined with passive scalar tracking, the impact of flow features transporting and mixing a scalar, such as a potentially infectious aerosol, can be deciphered.

This work presents CFD simulations of the airflow in a mock isolation room at the National Institute for Occupational Safety and Health (NIOSH) following the room, patient and healthcare worker (HCW) dimensions specified by Thatiparti et al. (2016a; 2016b). In Figure 1, the flow streamlines obtained from one of these simulations are visualized, particularly the streamlines emanating from the patient's mouth. The streamlines are generated by tracking air particles emanating from the region near the patient's mouth that serve to trace the flow pathlines over time. As can be seen in Figure 1, the flow is characterized by a short-circuiting or channeling between the patient's mouth and the exhaust air vent, creating an uninterrupted path that would enable aerosols discharged from the patient's mouth to exit the room quickly without causing great exposure to the HCW. Such short-circuiting and its desirable impact on HCW exposure has been shown by Memarzadeh and Xu (2012) in their CFD simulations. However, note that short-circuiting is accompanied by dead zones or recirculation zones (seen for example at the corners of the room in Figure 1b) which can accumulate aerosols for extended periods. Short-circuiting and dead zones can be viewed from a transportation analogy where short circuits are analogous to highways and dead zones are analogous to roundabouts. A fluid parcel caught in a short circuit will quickly reach its final destination (the exhaust) while a fluid parcel caught in a dead zone will not be able to reach its final destination until it leaves the dead zone through the action of turbulent diffusion (i.e. spreading due to the action of irregular (turbulent) fluctuations in the fluid parcel paths).

A popular model for estimating the time it takes for aerosols expelled by a patient to exit a ventilated enclosed room is based on the classical completely mixed reactor assumption. This theoretical model assumes that the aerosol is instantaneously perfectly mixed throughout the room, proceeding to begin exiting the room immediately after it is expelled by the patient. The model predicts an exponential decay rate of the aerosol concentration in the room with decay rate equal to the room air changes per hour (ACH), which is the inverse of the theoretical mean residence time defined as $\tau$ = V/Q, where V is the volume of the room and Q is the air flow rate entering/exiting the room (i.e. the ventilation rate): $C(t) = C(0)\exp(-t/\tau)$. In the previous expression $C(t)$ is the average room concentration of aerosol as a function of time. Based on the earlier introductory results of Figure 1, mixing conditions within hospital rooms are far from perfect, as a result of short-circuiting and dead zones, thus the previous theoretical model based on the perfectly mixed assumption can lead to errors. Indeed, Memarzadeh and Xu (2012) showed that increased short-circuiting between the patient and the exhaust air vents in an isolation room lowers exposure of a HCW to infectious aerosols, an effect that cannot be represented by the theoretical model.

The finding of Memarzadeh and Xu (2012) of the beneficial effect that short-circuiting can have on HCW exposure was obtained from CFD modeling over just the first five minutes of HCW exposure time, and is in agreement with the present results in Figure 1 (to be explained in more detail further below). Given the risks to clinicians working with virally infected patients, it is important to explore these findings beyond the initial 5 minutes of simulated exposure time, which the present work seeks to do. As such, one of the goals of the present study is to (1) investigate HCW exposure to aerosol discharged by a patient as a function of ACH in the isolation room for times greater than 5 minutes. Other goals are (2) to assess the relationship between short-circuiting (between the patient and the exhaust) and HCW exposure to aerosols discharged by the patient, and (3) to assess the efficacy of the perfectly mixed assumption traditionally employed for predicting HCW exposure.

The goals of the present study will be enabled by residence time distribution (RTD) analysis, a popular technique employed in the reactor modeling community, to be extended to the current application. In the classical approach to RTD analysis, a passive tracer (i.e. a scalar) pulse is released at the inlet of a stirred tank reactor (e.g. see sketch in Figure 2) and the tracer concentration is measured at the outlet over time. The time series of the tracer concentration at the outlet yields the so-called RTD curve which can be used to obtain residence time characteristics of the tracer in the reactor, such as mean residence time (MRT), and can be used to quantify the importance of short-circuiting and dead zone flow features (Figure 2) on the transport of tracer throughout the reactor. In the case of the isolation room studied in the present manuscript, RTD analysis is performed by releasing a tracer pulse from the mouth of the patient, similar to what a cough would do. Furthermore, the time series of the tracer concentration is recorded at the exhaust air vent to obtain the RTD curve.

# METHODS

## Reynolds-averaged flow equations

The core of CFD consists of solving the fluid flow equations numerically. In this study, the equations of fluid motion consist of the compressible Reynolds-averaged continuity and the Reynolds-averaged Navier-Stokes (RANS) equations. These two equations are given using indicial notation in (1) and (2), respectively:

$$\frac{\partial \rho}{\partial t} + \frac{\partial}{\partial x_i}(\rho u_i) = 0 \qquad (1)$$

$$\frac{\partial}{\partial t}(\rho u_i) + \frac{\partial}{\partial x_j}(\rho u_i u_j) = -\frac{\partial p}{\partial x_i} - \frac{\partial \rho \langle u_i' u_j' \rangle}{\partial x_j} + \frac{\partial}{\partial x_j}\left(\mu\left(\frac{\partial u_i}{\partial x_j} + \frac{\partial u_j}{\partial x_i} - \frac{2}{3}\delta_{ij}\frac{\partial u_l}{\partial x_l}\right)\right) \qquad (2)$$

where $\rho$ is the fluid (air) mean density; $p$ is mean pressure; $\mu$ is dynamic molecular viscosity; $u_i$ is the mean velocity in the i-th direction; $x_i$ is the coordinate in the i-th direction (the $x$, $y$, or $z$ direction) in Cartesian components; $t$ is time. The Reynolds (or turbulent) stress in (2), defined as $-\rho \langle u_i' u_j' \rangle$, is induced by the turbulent nature of the flow field, and is defined in terms of the ensemble average (denoted by the brackets) of the product of fluctuating (turbulent) velocity components $u_i'$. Note that as per the classical Reynolds decomposition, the total velocity is the mean velocity, $u_i$, plus the fluctuating velocity $u_i'$. Following the well-known RANS simulation methodology employed here, the mean velocity $u_i$ is computed or resolved via (1) and (2) while the Reynolds stress is modeled via the eddy (or turbulent) viscosity assumption

$$-\rho\langle u_i' u_j'\rangle = \mu_t \left(\frac{\partial u_i}{\partial x_j} + \frac{\partial u_j}{\partial x_i} - \frac{2}{3}\delta_{ij}\frac{\partial u_l}{\partial x_l}\right) - \frac{2}{3}\rho k \delta_{ij} \quad (3)$$

In order to calculate the eddy viscosity, $\mu_t$, and the turbulent kinetic energy, $k$, appearing in (3), the four-equation transition SST turbulence model is employed (Menter et al. 2004). This model is chosen for its ability to accurately capture low free stream turbulence environments as well as swirl flows which are important characteristics in ventilation flows, such as those considered here.

The flow mean temperature, $T$, is calculated via

$$\rho C_p \frac{\partial T}{\partial t} + \rho C_p \frac{\partial u_i T}{\partial x_i} = -\frac{\partial}{\partial x_i}\left(\kappa_e \frac{\partial T}{\partial x_i}\right) \quad (4)$$

Note that the mean density in (1)-(3) is obtained from the temperature through the ideal gas law. In (4), the effective conductivity $\kappa_e$ is the sum of molecular and turbulent conductivities, with the turbulent conductivity computed through the eddy viscosity obtained from the transition SST closure.

**Concentration transport equation**

The fluid is taken as a mixture composed of air and aerosol parcels with the tracer parcels having the same density as the air, $\rho$. Tracer mass transport is computed based on the convection-diffusion equation for tracer mass fraction:

$$\frac{\partial \rho Y}{\partial t} + \frac{\partial \rho u_i Y}{\partial x_i} = -\frac{\partial J_i}{\partial x_i} + \rho S \quad (5)$$

where $Y$ is the tracer mass fraction or concentration, $(1 - Y)$ is air mass fraction, $J_i$ the mass diffusion flux, and $S$ the tracer source.

The diffusion flux is defined as:

$$J_i = -\left(\rho D_m + \frac{\mu_t}{S_{c_t}}\right)\frac{\partial Y}{\partial x_i} - \frac{1}{T}\frac{\partial T}{\partial x_i} \quad (6)$$

where $D_m$ and $D_T$ are the mass diffusion coefficient and the thermal diffusion coefficient, respectively.

The transport of aerosol droplet nuclei corresponding to particles less than 20 μm in diameter has been shown to be well-modeled by the convection-diffusion equation in (5) (Liu and Zhai, 2007). Thus, the tracer modeled through equation (5) can be considered as aerosol droplet nuclei of this same size range.

To provide a source term, $S$, in equation (5), corresponding to a source of tracer, the concentration of aerosol is considered within the volume of a single cough from the patient over the amount of time it is required to produce the cough. In other words, this source term is kept on for an initial time corresponding to the period of a single cough, which is about 0.2 seconds, and then turned off for the remainder of the simulation.

## Computational Setup

The numerical simulations were performed for two hospital isolation room designs specified by Thatiparti et al. (2016a; 2016b). The first design considered was an optimal design that promotes short-circuiting between the patient and exhaust vent by placing the exhaust vent closer to the patient which can be seen in Figure 3a. A suboptimal design was also considered where the exhaust vent is located across the room from the patient as shown in Figure 3b, expected to lead to greater exposure of the HCW to aerosols expelled by the patient. The latter design provided the basis for the results presented in the Introduction.

Boundary conditions and material properties of surfaces in the room were assigned following Thatiparti et al. (2016a; 2016b) and are summarized in Tables 1-3.

Meshing of the computational domain was carried out using the cut-cell algorithm (Ingram et al., 2003), a tessellation-based Cartesian method, chosen for its ability to rapidly generate good quality meshes for simple geometries. Both room configurations maintain nearly the same number of mesh elements at around 3 million, significantly more than the 1.6 million mesh elements of Thatiparthi et al. (2016a; 2016b). Maximum skewness of mesh elements was kept to 0.7 (following recommendations in the literature, e.g. ANSYS Fluent User's Guide, 2016), aiding the stability and convergence of the solution.

The QUICK (Quadratic Upstream Interpolation for Convective Kinetics) scheme was used to discretize the momentum and temperature (energy) equations in (2) and (4), respectively, and the turbulence transition SST model transport equations (not shown). These equations were integrated in time arriving at a steady state solution for the flow and temperature distribution throughout the room. The corresponding pressure was obtained via a staggered-grid pressure interpolation scheme (similar to the classical schemes described by Patankar (1980)), ensuring satisfaction of the mass conservation equation in (1). Finally, the mass fraction equation in (5) was discretized via a $2^{nd}$ order implicit method in time and a $1^{st}$ order upwind method in space to obtain time dependent solutions.

The time averaged flow was initially solved in the absence of temperature (compressible) effects. Once the homogeneous flow was obtained, the temperature equation in (4) was activated allowing for secondary flow characteristics (such as natural convection) to be incorporated into the solution obtained from the computation of the primary flow without temperature effects. This improved the rate of convergence towards a steady solution of the system. With the coupled momentum and temperature equations solved, the scalar transport equation in (5) (with the source term (representing the patient's cough) turned on over the first 2 seconds), was run to track the aerosol concentration throughout the room and over time.

The time step for the scalar transport equation in (5) was chosen such that the Courant-Friedrichs-Levy (CFL) number (Patankar 1980) based on ACH was kept roughly constant for each case, to allow consistent comparison between the cases with different ACH. Simulations at 6 ACH and 12 ACH for suboptimal and optimal rooms designs were performed, yielding a total of 4 simulations to be analyzed. The time step for the 6 ACH and 12 ACH simulations was 0.125 s and 0.0625 s, respectively.

# RESULTS

## Aerosol Transport

Both room configurations ran with an ACH of 6 and 12, following Memarzadeh and Xu (2012) and Thatiparti (2016a; 2016b). Figures 4 and 5 correspond to the results for the 6 ACH and 12 ACH cases respectively. In Figures 4a,b and 5a,b, the normalized aerosol concentrations or mass fractions recorded at the exhaust for the optimal room design configuration show a large spike at early times due to the strong short-circuiting that exists between the patient's mouth and the exhaust. For later times, the concentration time series exhibits a behavior similar to that of perfectly mixed conditions for which the concentration decays exponentially. The optimal design cases show that the short-circuiting is much stronger for ACH = 12. Additionally, the suboptimal design cases show marginal short-circuiting compared to the optimal design, revealed by the significantly reduced peak in concentration time series in the suboptimal design cases.

In both of the optimal design cases, in addition to the large spike in aerosol concentration measured at the exhaust associated with strong short-circuiting, there are notable secondary peaks or waves of aerosol registered at the exhaust proceeding the large spike (Figs. 4a,b and

5a,b). For example, in Fig. 4a,b, two secondary peaks can be observed at times t ~ 0.9 min. and 1.5 min. These secondary peaks are attributed to the impact of dead zones, which serve to initially retain aerosol for a limited time, with the aerosol eventually escaping the dead zones as a result of turbulent dispersion.

In Figures 4c and 5c, considering the normalized bulk concentration (the normalized concentration averaged over the volume of the room), once again there is notable difference between the 6 and 12 ACH cases. The low ACH case for the suboptimal design shows higher bulk concentration at early times but overall, it is very close to the concentration obtained with the theoretical model based on the perfectly mixed assumption. In contrast, the high ACH case for the suboptimal design shows that the bulk concentration is higher than the concentration obtained with the theoretical model. And the optimal design case provides a much more pronounced reduction of the bulk concentration.

## HCW exposure

The HCW exposure, which is the integral of the average aerosol mass fraction recorded near the HCW over time, is considered in Figures 4d and 5d. The HCW exposure calculated is based on the aerosol mass fraction averaged over the region within 0.3 meters of the head of the HCW.

Over early times, less than one minute, the suboptimal design shows negligible HCW exposure while the optimal design shows immediate exposure. However, for later times, the opposite is true. For example, for times less than 2 minutes, the 12 ACH suboptimal design case provides less HCW exposure, but as time progresses it quickly surpasses the HCW exposure in the optimal design case and the exposure predicted by the theoretical model by a significant margin. Considering this, if a limit were to be set for an amount of exposure for the HCW, it can be seen that in the optimal design case with ACH = 12, the amount of time a HCW can remain in the room would be far longer than that in the suboptimal design case, as well as that predicted by the theoretical model based on the perfectly mixed reactor assumption.

The optimal design shows significant reduction in HCW exposure compared to the suboptimal design. This holds for both the 6 and 12 ACH cases, but it is more notable in the 12 ACH case. Furthermore, for the 12 ACH case, the theoretical model based on the perfectly mixed assumption predicts nearly double the amount of exposure for large times as that recorded in the optimal design case. When considering the 6 ACH cases, the theoretical model yields a HCW exposure relatively close to that measured in both the optimal and suboptimal design cases. Therefore, the perfectly mixed assumption is more suitable for 6 ACH. However, this assumption does not hold as well for 12 ACH, as there is significant discrepancy between both the suboptimal and optimal designs and the theoretical model based on the perfectly mixed assumption.

In Figure 6, it is seen that a greater ACH does not necessarily reduce exposure of the HCW to patient-expelled aerosols for times less than 5 minutes. This can be concluded by comparing the curves for the 6 ACH and 12 ACH cases for the suboptimal room configuration, which are close to each other for times between t = 2 and 5 minutes (shaded time range in Fig. 6). Instead, the controlling factor in reducing exposure over this time period is the optimal placing of the exhaust relative to the patient, as originally found by Memarzadeh and Xu (2012). However, this is not the case for exposure times greater than 5 minutes, as the 12 ACH cases are seen to reduce HCW exposure for both optimal and suboptimal room designs, relative to the corresponding 6 ACH cases.

## RTD analysis

RTD analysis is based on the time series of the aerosol or scalar concentration recorded at the exhaust in Figures 4a,b and 5a,b, or $C(t)$. The RTD curve is defined as the normalized $C(t)$

$$E(t) = \frac{C(t)}{\int_0^\infty C(t)\, dt} \tag{7}$$

with the first moment of the RTD providing the mean residence time

$$\tau = \int_0^\infty t\, E(t)\, dt \tag{8}$$

(Fogler, 2018). As noted in the introduction, the theoretical MRT is the ratio of the volume of the room to the volumetric air flow rate entering and exiting the room, $\tau_{theor} = V/Q$. The theoretical MRT corresponds to perfectly mixed conditions, however, as highlighted in the previous sub-section, mixing conditions within an isolation room deviate from perfect due to the presence of short-circuiting and dead zones. The strengths or influence of these flow features determine how much greater or lower the actual MRT is compared to theoretical MRT. For example, strong short-circuiting implies that the tracer will exit the room rapidly, leading to a lower actual MRT than the theoretical MRT. Weaker short circuiting and stronger dead zones imply that the dead zones will be able to retain elevated tracer concentrations for extended periods of time, resulting in a greater actual MRT than the theoretical MRT. Overall, the expression in (8) provides the actual MRT under the influence of the true mixing conditions that determine the RTD. As such, the CFD-computed RTD can be used to obtain the actual simulated MRT, to be referred to as $\tau_{CFD}$ in the discussions below.

The RTD can provide further insight into the nature of the flow in the system. The cumulative RTD defined as

$$F(t) = \int_0^t E(t)\, dt \qquad (9)$$

(Fogler, 2018) can be used to calculate the time it takes a certain percentage of the aerosol to exit the room, as depicted in Figure 7. For example, from the sketch of Figure 7 and the definition in (9), it can be seen that $t_{10}$ refers to the time it takes for 10% of the aerosol emitted during the patient's single cough event to exit the room (i.e. $t_{10}$ corresponds to the time at which $F = 0.1$). Similarly, $t_{90}$ refers to the time it takes for 90% of the aerosol emitted to exit the room.

As noted by Wilson and Venayagamoorthy (2010), the shape of the cumulative RTD curve is indicative of the type of flow in the system. For example, a steeper gradient in the $F(t)$ curve corresponds to flow conditions closer to plug flow, which is dominated by advective transport relative to diffusion. A flow characterized by strong short-circuiting would give rise to a steeper gradient in $F(t)$, due to the strong advection (flow transport) characterizing the short-circuit. A flatter gradient in $F(t)$ is indicative of a flow dominated by diffusion relative to advective transport. The latter would correspond to a flow tending towards a perfectly mixed system.

The quantity $(t_{90} - t_{10})$, as measured from $F(t)$, can been used as an index to help characterize the shape of $F(t)$, and thus the flow condition. A larger value of $(t_{90} - t_{10})$ would correspond to a flatter gradient in $F(t)$ and thus a flow dominated by diffusion, while a smaller value would correspond to a steeper gradient and thus a flow dominated by advection. The quantity $t_{10}$ on its own can be used as a measure of short-circuiting. A smaller value of $t_{10}$ would correspond to stronger short-circuiting and vice-versa (Teixeira and Siqueira 2008).

Figure 8 and Table 4 present the cumulative RTD and list several residence time characteristic indices (such as $\tau_{CFD}$, $t_{10}$, and $t_{90} - t_{10}$), respectively, for the aerosol (tracer) released by the patient simulated via CFD. Also included are results based on the assumption that the aerosol is perfectly mixed throughout the isolation room.

In Figure 8b, the cumulative RTD ($F(t)$) for the 12 ACH optimal design case is characterized by a sharper gradient than for the other cases through the first minute of simulation. This is indicative of the stronger advection relative to diffusion, the former induced by the stronger short-circuiting caused by the optimal placing of the exhaust. This trend can also be seen for the 6 ACH cases (Figure 8a), but it is not as evident. For the most part, the $F(t)$ for the 6 ACH optimal and suboptimal cases follow the trend of the theoretical model based on the perfectly mixed assumption (Figure 8a). Overall, the trends observed in Figure 8 in terms of the gradient of $F(t)$ are quantified in terms of the values of $(t_{90} - t_{10})$ listed in Table 4.

The impact of short-circuiting, and thus the optimal placing of the exhaust, can also be observed in terms of MRT, especially for the 12 ACH cases. For these cases, the increased short-circuiting resulting from the optimal placing of the exhaust causes the CFD-computed MRT ($\tau_{CFD}$) to drop by 52% (see Table 4). Meanwhile, for the 6 ACH cases, the corresponding drop is 6.6% (Table 4). The impact of short-circuiting can also be quantified by comparing the values of $\tau_{CFD}$ with the corresponding values of $\tau_{theor}$, listed in Table 4. Note that for a fixed value of ACH, the values of $\tau_{theor}$ are equal for optimal and suboptimal designs, as in the theoretical model (based on the perfectly mixed assumption) the aerosol is distributed homogenously throughout the room without accounting for heterogeneities in the flow field caused by short-circuiting and dead zones. In Table 4, it can be seen that for the optimal design cases, the values of $\tau_{CFD}$ are lower than the values of $\tau_{theor}$, as a result of the short-circuiting able to rapidly expel a significant portion of the aerosol emitted by the patient. Note that ratios $\tau_{theor}/\tau_{CFD}$ for the 12 ACH and 6 ACH cases are 1.9 and 1.1, respectively, indicative of the stronger short-circuiting induced by the greater ACH. Furthermore, in Table 4, it can be seen that for the suboptimal design cases, the values of $\tau_{CFD}$ are greater than the values of $\tau_{theor}$, as a result of the short-circuiting not being strong enough to rapidly expel the aerosol, allowing the dead zones to retain elevated aerosol concentrations. For the suboptimal cases in which dead zones are the dominant flow features (rather than short-circuiting), a higher ACH can still reduce the aerosol MRT through greater levels of diffusion that allow the aerosol to escape dead zones at a faster rate.

The strength of short-circuiting for the different cases is further considered in Table 4 in terms of $t_{10}$. Based on the ratio of $t_{10}$ between the 6 ACH cases, it can be concluded that short-circuiting in the optimal design case is about 2.6 times stronger than in the suboptimal design case. Similarly, for the 12 ACH cases, short-circuiting in the optimal design case is about 8.4 times stronger than in the suboptimal design case. This is consistent with the ratio of the concentration peaks between the optimal and suboptimal design cases for 6 ACH (Fig. 4a) and 12 ACH (Fig. 5a), and with the earlier conclusion that the impact of short-circuiting is more evident for the 12 ACH cases than for the 6 ACH cases.

**HCW exposure and residence time indices**

In water and wastewater treatment, hydraulic efficiency (measured in terms of residence time indices) is often tied to treatment efficiency. For example, the United States Environmental Protection Agency (USEPA, 2003) has established the baffle factor (BF), defined as $BF = t_{10}/\tau_{theor}$, in order to evaluate the disinfection efficiency of ozonation tanks in municipal drinking water treatment plants. An analogous approach is taken here, but with the goal of assessing the HCW exposure in terms of residence time characteristics.

Table 4 lists long-term HCW exposure for the cases simulated, where the long-term exposure is defined as the asymptotic value of the HCW exposure plotted in Figures 4d and 5d. Table 4 also lists short-term HCW exposure, defined here as the HCW registered at time t = 3 minutes. Note that at t = 3 mins., the HCW exposure for the 6 ACH suboptimal and optimal design cases and for the 12 ACH suboptimal design case are nearly identical as seen in Figure 6. Meanwhile, a lower short-term exposure is only observed for the 12 ACH optimal design case. Thus, as described earlier in sub-section 3.2, a greater ACH does not necessarily reduce short-term exposure of the HCW to patient-expelled aerosols. Instead, the controlling factor in reducing short-term exposure is the optimal placing of the exhaust relative to the patient, as originally found by Memarzadeh and Xu (2012).

Looking at the various residence time indices listed in Table 4 for the cases simulated, the trend of CFD-predicted MRT, $\tau_{CFD}$, follows closely to that of the long-term HCW exposure. Furthermore, the trend of $(t_{90} - t_{10})/\tau_{theor}$ approximates that of the short-term HCW exposure. This latter trend is corroborated by Figure 9, which shows that the cumulative RTDs vs. time scaled by $\tau_{theor}$ for the 6 ACH cases with suboptimal and optimal designs and for the 12 ACH case with suboptimal design follow close to each other while the cumulative RTD for 12 ACH case with optimal design deviates significantly from these others.

Panels (a) and (b) of Figure 10 show the trends between residence time indices and HCW exposure by plotting long-term HCW exposure vs. $\tau_{CFD}$ and short-term HCW exposure vs. $(t_{90} - t_{10})/\tau_{theor}$, respectively, as listed in Table 4. Linear best fits representing these trends are also included in Figure 10 with "goodness-of-fit" measures of $R^2 > 0.92$. In particular, $R^2$ for the linear best fit of long-term HCW exposure vs. $\tau_{CFD}$ is 0.99. Such a close linear relationship between MRT and long-term exposure across all cases simulated demonstrates that the long-term

HCW exposure depends on both short-circuiting and dead zones (and not short-circuiting alone) as these are the key factors determining the MRT. Meanwhile, the strong linear relationship between $(t_{90} - t_{10})/\tau_{theor}$ and short-term HCW exposure points to the significance of the strength of short-circuiting (i.e. advection) relative to diffusion in determining the short-term HCW exposure. In other words, the fact that neither $t_{10}$ or $t_{10}/\tau_{theor}$ is correlated with the short-term HCW (as can be concluded from Table 9) points to the fact that short-term HCW exposure is not purely dependent on the strength of the short-circuiting, but rather on the strength of short-circuiting relative to diffusion (i.e. on the index $(t_{90} - t_{10})/\tau_{theor}$).

Based on the previous results, the following question may be addressed: Why is promoting short-circuiting via the optimal design effective in lowering short-term HCW exposure with ACH = 12, but not with ACH = 6? In the case with ACH = 12, given the high circulation rate (i.e. high flow velocities) afforded by the high value of ACH, the resulting short-circuit is able to lower the aerosol concentration rapidly, thereby reducing the importance of aerosol diffusion throughout the room. In other words, the strengthening of short-circuiting in switching from the suboptimal to the optimal design comes at the expense of diffusion, thus reducing the value of $(t_{90} - t_{10})$ as seen in Table 4. In the case with ACH = 6, the lower circulation rate (i.e. the lower flow velocities) limits the strengthening of the short-circuiting when switching from the suboptimal to the optimal design (relative the case with ACH = 12). This can be seen in Table 4 through the less significant drop in $t_{10}$ in the ACH = 6 case when switching from the suboptimal to the optimal design compared to the corresponding drop in the ACH = 12. A milder strengthening of the short-circuit in the ACH = 6 cases (compared to the ACH=12 case) is responsible for leaving higher levels of aerosol available for spreading through diffusion, such that the strength of the short-circuit relative to the strength of diffusion remains nearly the same when switching from the suboptimal to the optimal case. Note that the milder short-circuiting in the ACH = 6 optimal design case compared to the ACH = 12 optimal design case allows for diffusion to remain important by allowing higher aerosol concentrations to be spread by diffusion. Also note that these results show that the strengthening of the short-circuit (through changes in the placement of the exhaust) and its lowering of the short-term HCW exposure is ultimately limited or capped by the amount of ACH.

## CONCLUSIONS

CFD simulations of aerosol dispersion in a mock hospital isolation room were presented at ACH = 6 and ACH = 12. The goals were (1) to investigate HCW exposure to aerosol discharged by a patient as a function of ACH in the isolation room for times greater than 5 minutes, (2) to assess the relationship between short-circuiting (between the patient and the exhaust) and HCW exposure to aerosols discharged by the patient, and (3) to assess the efficacy of the perfectly mixed assumption traditionally employed for predicting HCW exposure. RTD analysis of the CFD simulation results led to the following main findings addressing the goals:

(i) It was seen that the effectiveness of short-circuiting between the patient and exhaust at lowering HCW exposure is dramatically affected by positioning of the exhaust for high ACH, but this effect is less pronounced for low ACH.

(ii) It was observed that a high ACH reduces HCW exposure over periods greater than about 6 minutes even for suboptimal short-circuiting designs, but not necessarily over shorter periods (less than 5 minutes). Memarzadeh and Xu (2012) reported that optimal placement of the exhaust relative to the location of the patient is more important in reducing HCW exposure than increasing ACH. The present simulations confirm this, but this importance is only seen approximately between the first 2 and 5 minutes of exposure, as for periods of exposure longer than about 6 minutes, a higher ACH does prove beneficial even for a suboptimal room design.

(iii) The theoretical model based on the perfectly mixed assumption works relatively well for predicting HCW exposure and bulk (room averaged) aerosol concentration at low ACH, however, it does not work as well for these quantities at high ACH. Furthermore, a suboptimal exhaust position tends to bring about HCW exposures greater than the long-term exposure predicted via the theoretical model, while an optimal exhaust position results in less long-term exposure than the theoretical model, regardless of ACH. The former was attributed to the fact that for a suboptimal room design, a greater amount of aerosol becomes trapped in dead zones leading to greater HCW exposure than if the aerosol was perfectly mixed. The latter was attributed to the fact that a stronger short-circuiting in the optimal room design is able to rapidly expel a significant amount of aerosol from the room, thereby reducing the HCW exposure compared to if the aerosol was perfectly mixed condition throughout the room.

(iv) RTD analysis showed that MRT ($\tau$) is well-correlated with long-term (> 6 mins.) HCW exposure, while the index $(t_{90} - t_{10})/\tau_{theor}$ is correlated with short-term (< 5 mins.) HCW exposure. Due to its close connection with MRT, long-term HCW exposure was thus shown to be dictated by short-circuiting and dead zones. Stronger short-circuiting via optimal exhaust placement and higher diffusion via higher ACH (thereby reducing the impact of dead zones) can both independently reduce MRT and thus long-term HCW exposure. Furthermore, due to its close connection with $(t_{90} -$

$t_{10})/\tau_{theor}$, short-term HCW was thus shown to be dictated by the strength of short-circuiting relative to diffusion (and not just purely by short-circuiting).

RTD analysis has been a staple of drinking water treatment plant design as a way to ensure disinfection standards set by the U.S. EPA. The closed connection observed between residence time indices and HCW exposure, summarized in item (iv) above, suggests that future designs of hospital isolation rooms could benefit from RTD analysis, either conducted via CFD as done here, or via physical experiments, without the need to employ a HCW proxy and measure their exposure directly. This would be analogous to the standard practice of employing RTD analysis at the design stage of drinking water disinfection tanks to ascertain the expected disinfection efficiency of the tank without performing direct measurements of contaminant levels.

## ACKNOWLEDGMENTS

The authors would like to acknowledge the use of the services provided by Research Computing at the University of South Florida.

## FUNDING STATEMENT

This research was partially supported through the USF COVID-19 Rapid Response Grant Program.

## DATA AVAILABILITY STATEMENT

The data that support the findings of this study are available from the corresponding author, AET-M, upon reasonable request.

# FIGURES

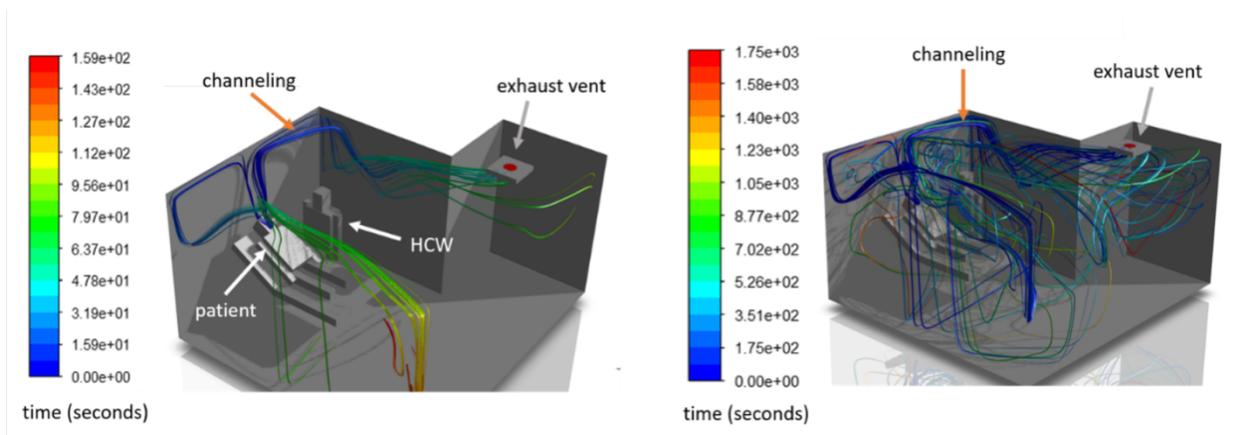

**Figure 1.** [Flow path lines or streamlines emanating from the mouth of the patient in a CFD simulation of a mock isolation room at the NIOSH. The simulation was made following the room and object dimensions, and thermal and ventilation specifications (with air changes per hour, ACH=6) given by Thatiparti et al. (2016a; 2016b) (to be given in detail further below). The streamlines are color-coded by time. The fewer streamlines on the left correspond to the less time allowed for the air particles emanating from the patient's mouth to trace out the pathlines].

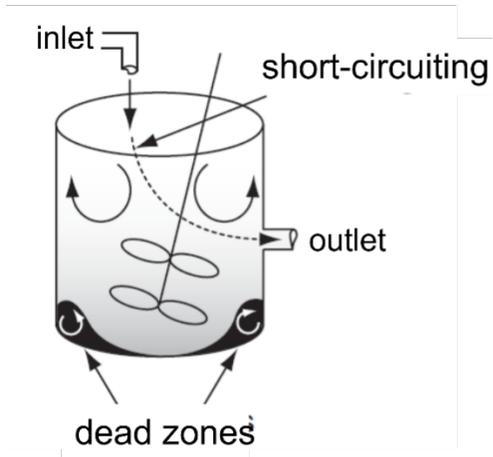

**Figure 2.** [Classic illustration of flow dead zones and inflow-outflow short-circuiting coexisting with intense mixing caused by a mechanical agitator in a stirred tank reactor].

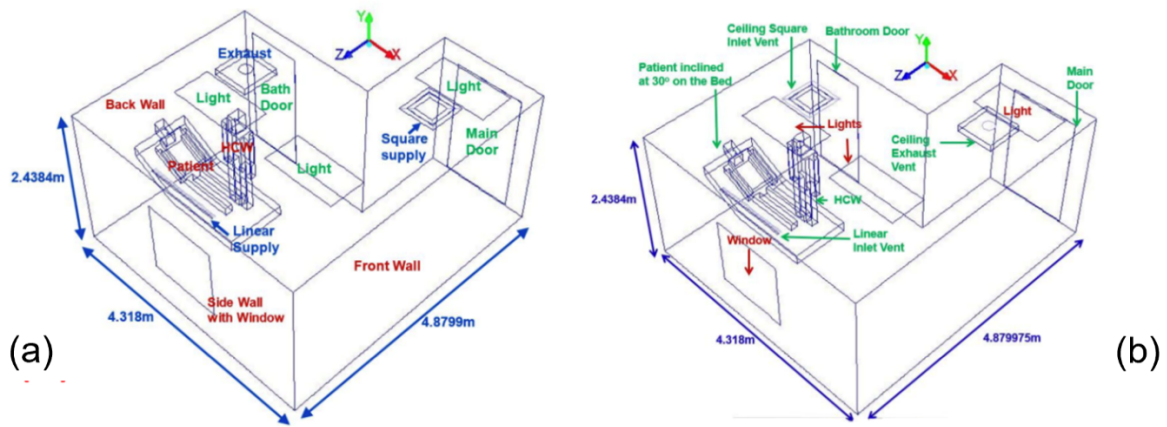

**Figure 3.** [Two hospital isolation room designs, given by Thatiparti et al. (2016a; 2016b), considered in the present CFD analysis: an optimal design that promotes channeling between the patient and exhaust vent shown in (a) and a suboptimal design in (b)].

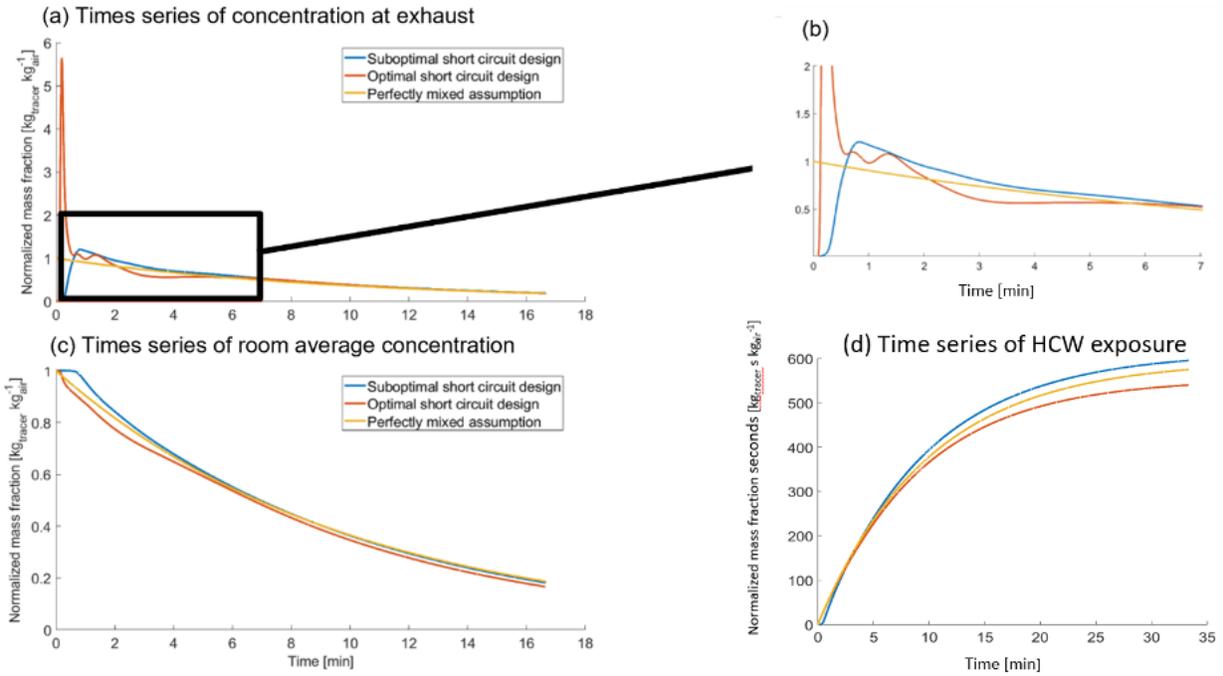

**Figure 4.** [Aerosol (tracer) concentration measured at the exhaust in suboptimal and optimal design cases, and for the corresponding theoretical model based on the perfectly mixed assumption cases for ACH = 6].

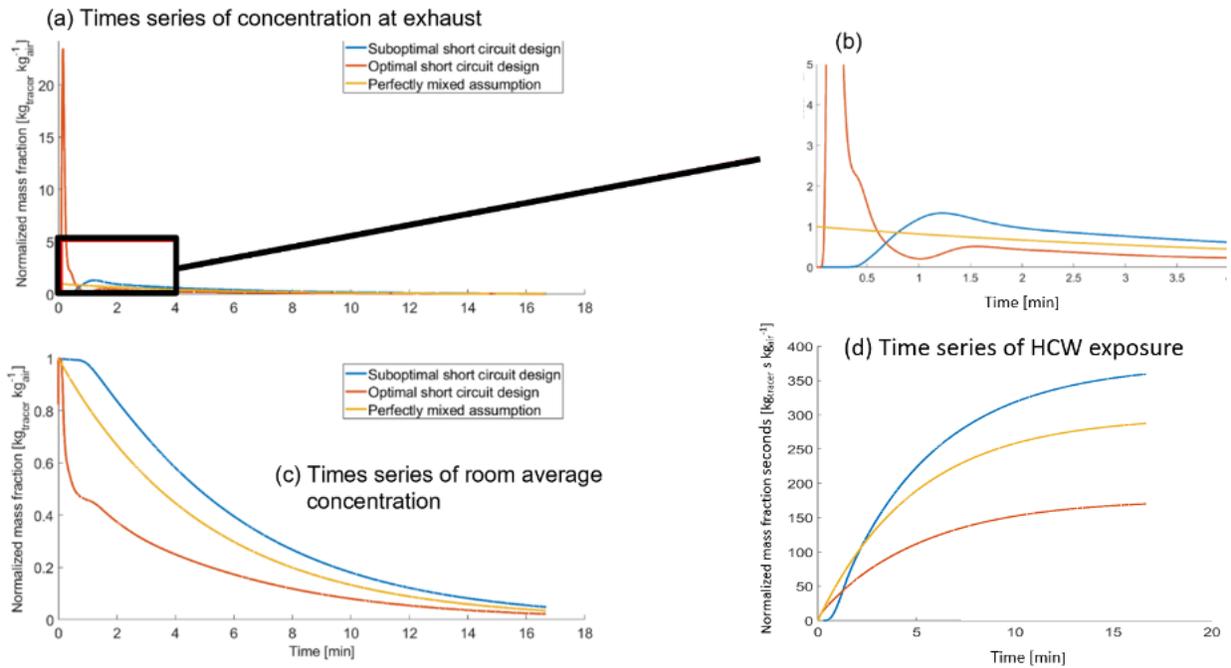

**Figure 5.** [Aerosol (tracer) concentration measured at the exhaust in suboptimal and optimal design cases, and for the corresponding theoretical model based on the perfectly mixed assumption cases for ACH = 12].

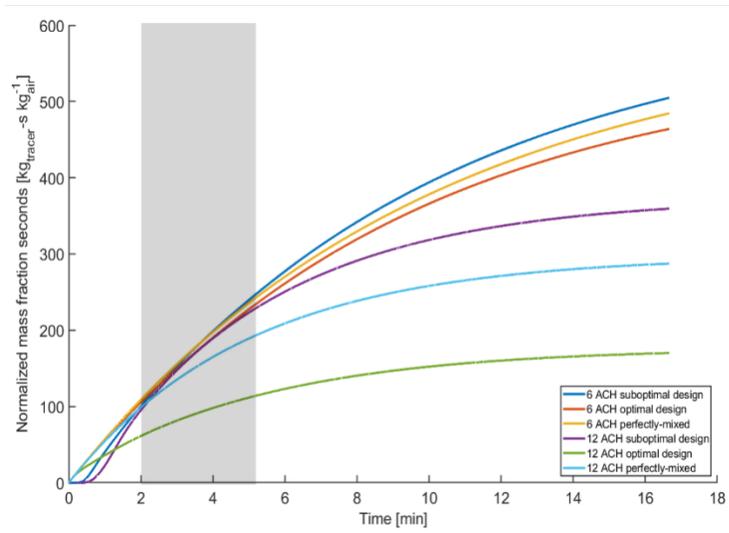

**Figure 6.** [HCW exposure for suboptimal, optimal, and perfectly mixed cases at 6 and 12 ACH].

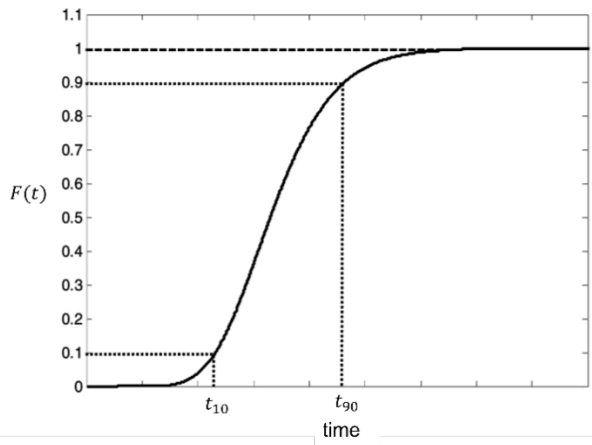

**Figure 7.** [Sketch of cumulative RTD function, $F(t)$, for an arbitrary system (adapted from Wilson and Venayagamoorthy, 2010)].

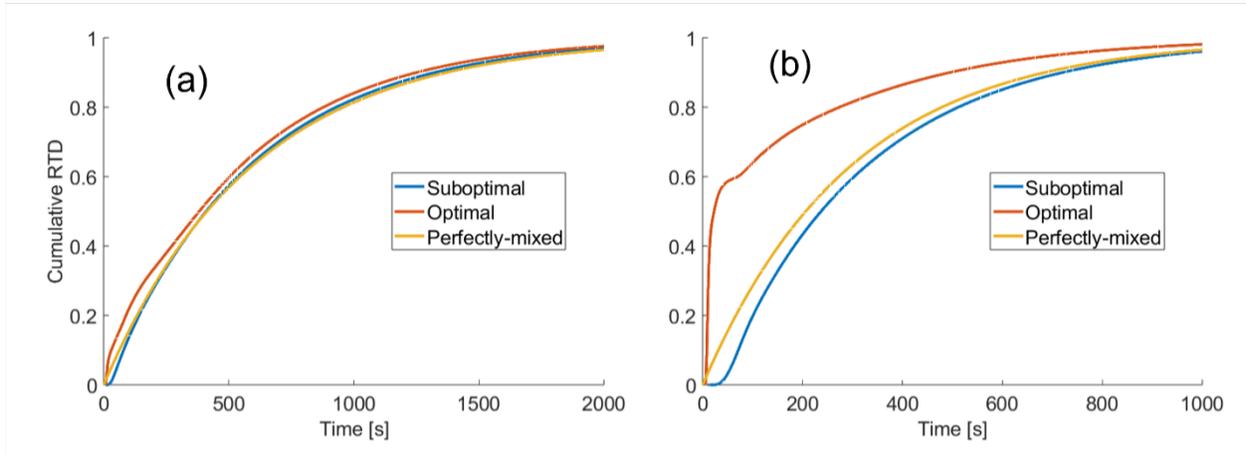

**Figure 8.** [Cumulative RTD function, $F(t)$, for (a) 6 ACH and (b) 12 ACH cases].

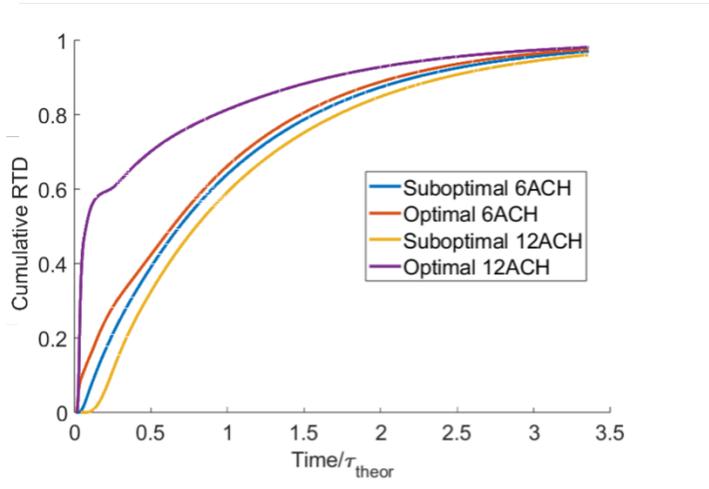

**Figure 9.** [Cumulative RTD function, $F(t)$, for 6 ACH and 12 ACH cases vs. time with time scaled by theoretical MRT, $\tau_{theor}$].

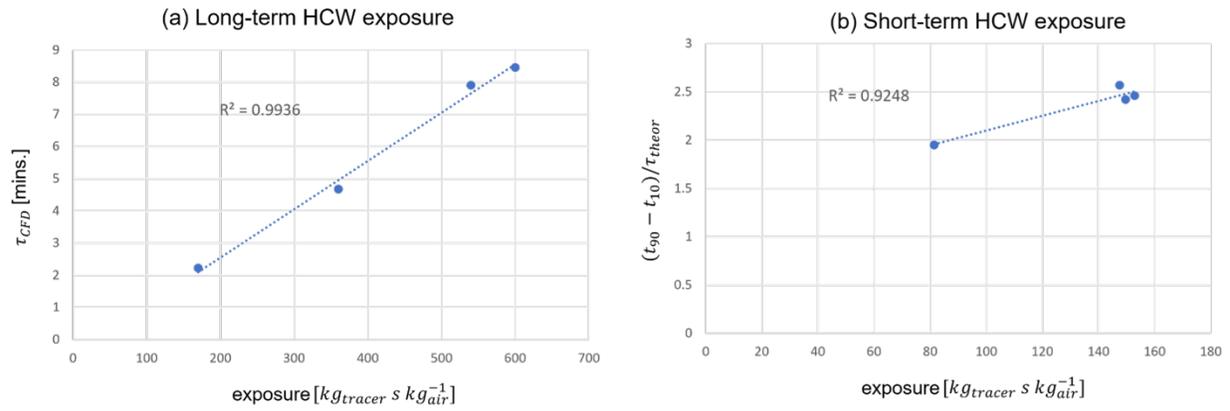

**Figure 10.** [(a) Long-term HCW exposure vs. $\tau_{CFD}$ and (b) short-term HCW exposure vs. $(t_{90} - t_{10})/\tau_{theor}$ data points listed in Table 1 and corresponding linear best fits].

# TABLES

**Table 1.** [Material properties of surfaces].

| Surface | Material | Conductivity [W m$^{-1}$ K$^{-1}$] | Specific Heat [J kg$^{-1}$ K$^{-1}$] | Density [kg m$^{-3}$] |
|---|---|---|---|---|
| Window | Window glass | 0.96 | 840 | 2500 |
| Patient/HCW | Human skin | 0.206 | 3558 | 1027 |
| Overhead lights | Plastic | 1.005 | 1670 | 1250 |
| Doors | Wood | 0.173 | 2310 | 700 |
| Room walls | Dry wall | 3.1 | 1090 | 600 |
| Patient bed | Cotton fabric | 0.043 | 1162 | 1540 |
| Floor | Vinyl flooring | 0.28 | 900 | 1450 |

**Table 2.** [Momentum and temperature equations boundary conditions (BC). All modeled surfaces are no-slip with a specified temperature as listed. Zero total (advective plus diffusive) normal flux was assigned for the scalar mass fraction at all of the surfaces listed, except at the exhaust (return housing) where only the diffusive normal flux was prescribed as zero].

| Surface | Temperature [K] | Material |
|---|---|---|
| Patient | 312 | Cotton |
| Patient Head | 312 | Skin |
| HCW | 309 | Cotton |
| HCW Head | 309 | Skin |
| Bed | 294 | Cotton |
| Lights | 294 | Plastic |
| Walls | 294 | Dry Wall |
| Window | 294 | Glass |
| Floor | 294 | Vinyl Flooring |
| Doors | 294 | Wood |
| Ceiling | 294 | Dry Wall |
| Supply/Return Housing | 294 | Aluminum |

**Table 3.** [Inflow boundary conditions (BC). Note that both linear and square air supplies direct part the air flow at 45 degree angles from the ceiling. See Thatiparti et al. (2016a; 2016b) for more details about the geometry].

| ACH | Surface | Mass flow rate [kg/s] |
|---|---|---|
| 6 | Linear Supply | 0.0462 |
| 6 | Square Supply | 0.0377 |
| 12 | Linear Supply | 0.0923 |
| 12 | Square Supply | 0.0755 |

**Table 4.** [Residence time indices and HCW exposures]

| Case | $\tau_{theor}$ [mins.] | $\tau_{CFD}$ [mins.] | $t_{10}$ [mins.] | $(t_{90} - t_{10})$ [mins.] | $\dfrac{(t_{90} - t_{10})}{\tau_{theor}}$ | long-term HCW exposure $\left[\dfrac{kg_{tracer} \cdot s}{kg_{air}}\right]$ | short-term HCW exposure $\left[\dfrac{kg_{tracer} \cdot s}{kg_{air}}\right]$ |
|---|---|---|---|---|---|---|---|
| 6 ACH suboptimal | 8.42 | 8.46 | 1.33 | 20.7 | 2.46 | 600 | 152.9 |
| 6 ACH optimal | 8.42 | 7.90 | 0.505 | 20.4 | 2.42 | 540 | 149.6 |
| 12 ACH suboptimal | 4.21 | 4.67 | 1.19 | 10.8 | 2.57 | 360 | 147.5 |
| 12 ACH optimal | 4.21 | 2.22 | 0.141 | 8.12 | 1.95 | 170 | 81.36 |